\documentclass[showpacs,prl,twocolumn,aps,superscriptaddress,preprintnumbers]{revtex4}
\usepackage{epsfig}
\usepackage{graphicx}
\usepackage{amsmath}

\begin{document}
\preprint{BNL-98513-2012-JA, RBRC-968}
\title{Anisotropy of photon production: Initial eccentricity or magnetic field.}
\author{Adam Bzdak}
\email[E-Mail:]{ABzdak@bnl.gov}
\affiliation{RIKEN BNL Research Center, Brookhaven National Laboratory, Upton, NY 11973,
USA}
\author{Vladimir Skokov}
\email{VSkokov@bnl.gov}
\affiliation{Department of Physics, Brookhaven National Laboratory, Upton, NY
11973, USA}
\date{\today }

\begin{abstract}
Recent measurements of the azimuthal anisotropy of direct photons in 
heavy-ion collisions at the energies of RHIC show that it is of the same order as
the hadronic one. This finding appears to  contradict the expected dominance
of photon production from a quark-gluon plasma at an early stage of a heavy-ion collision.
A possible explanation of the strong azimuthal anisotropy of the photons, given recently,
is based on the presence of a large
magnetic field in the early phase of a collision. In this letter,
we propose a method to experimentally  measure the
degree to which a magnetic field in heavy-ion collisions is
responsible for the observed anisotropy of photon production. 
The experimental test proposed in this letter may potentially change our understanding of the non-equilibrium 
stage and possible thermalization in heavy-ion collisions.

\end{abstract}

\maketitle

In contrast to hadronic observables, which strongly are influenced by the 
final state of the fireball created in a heavy-ion collision, 
the ``direct'' photons leave the fireball 
almost without interacting with the medium~\cite{Feinberg:1976ua,Shuryak:1978ij,McLerran:1984ay},
and thus they may play an important role in unraveling the properties of hot and dense matter.
According to expectations, based on the large yield of thermal photons, 
photon production is believed to be dominated by the hottest phase, quark-gluon plasma (QGP) at the early 
stage of a
collision at the top energies in RHIC (Brookhaven) and the LHC (CERN). 
Measurements from the PHENIX collaboration showed that the observed temperature of photon radiation 
at energy $\sqrt{s}=200$ GeV in heavy-ion collisions is about $T_{\rm ave} \approx 220$ MeV~\cite{Adare:2008ab}.
This value can be considered as an average over the entire evolution of the matter created in heavy-ion collisions. It is higher than the temperature of the phase transition   
and thus it supports the picture of photon production from the quark-gluon stage of the collision.
An alternative explanation of the apparent high temperature of the photon source
is the formation of  prethermal glasma shining photons early  in a collision~\cite{Chiu:2012ij}. 
In both scenarios, we  would expect that the photons' azimuthal anisotropy,
characterized by the second Fourier component
\begin{equation}
	  v^\gamma_2 (p_t) = \frac{\int d\phi \cos(2\phi) \frac{dN^\gamma}{dp_t d\phi}}{
		  \frac{dN^\gamma}{dp_t}},
\end{equation}
is  small~\cite{Chatterjee:2007xk}, because, at the early QGP stage, the evolving medium
does not develop an appreciable amount of azimuthal anisotropy. 
Others studied this phenomenon in hydrodynamic calculations that well  describe 
hadron production and hadron azimuthal anisotropy. The hydrodynamic calculations~\cite{Dion:2011pp},
as expected, demonstrated that the photon $v_2^\gamma$ is almost an order of magnitude smaller than the one of hadrons.
Taking into account fluctuating initial conditions or  viscosity does not significantly increase 
$v_2^\gamma$. 
However, recent
measurements by the PHENIX collaboration revealed that the anisotropy of
produced photons is very close to that of hadrons~\cite{Adare:2011zr}. 
This finding might be explained in several ways. 
For instance, van Hees and collaborators~\cite{vanHees:2011vb}  
investigated how the fireball's evolution is constrained by the photon
azimuthal anisotropy. To describe the photon spectra it was required
to introduce
a large acceleration of the fireball's expansion; this, however, is
not present in
conventional hydrodynamical calculations. 
Alternatively, we can assume 
that another unknown mechanism exists that is responsible for  photon anisotropy. Such
mechanism was recently proposed in Ref.~\cite{Basar:2012bp}. It  is based on the presence
of a large highly anisotropic magnetic field in heavy-ion collisions that  
could serve as a natural source of the photons' anisotropy. 
It is  maximal at the early stage of a collision, thereby  allowing us  to reconcile the large
thermal yield of photons with significant azimuthal anisotropy.  

The importance of the magnetic field was recognized in Ref.~\cite{Kharzeev:2007jp}, 
wherein the chiral magnetic effect~\cite{Kharzeev:2007jp,Fukushima:2008xe,Kharzeev:2009fn} was proposed and studied. 
The results
of Refs.~\cite{Kharzeev:2007jp,Skokov:2009qp,Bzdak:2011yy} showed that in
non-central heavy ion collisions the amplitude of magnetic field can reach
very high values up to a few $m_{\pi }^{2}\approx 10^{18}$ Gauss. This field
is large at the time of the collision and decreases inversely proportional to
the square of time \footnote{The produced medium response with a realistic values of electric conductivity may change this dependence only slightly. This problem will be addressed elsewhere.}. The magnetic field essentially is anisotropic and, on average,
points in the direction perpendicular to the reaction plane.  It also 
was demonstrated~\cite{Bzdak:2011yy} 
that event-by-event fluctuations of the magnetic (and electric) field
may play an important role on the level of relevant observables.

According to the calculations of several researchers~\cite{Basar:2012bp,Fukushima:2012fg,Tuchin:2012mf}, 
photons are emitted in the direction perpendicular to the magnetic field. Different microscopic mechanisms of
this emission were proposed recently, including the synchrotron
radiation of photons from moving charged quarks~\cite{Tuchin:2012mf}, 
and that reflecting  the scale anomaly of QCD$\times $QED~\cite{Basar:2012bp} 
and owing to the axial anomaly~\cite{Fukushima:2012fg}. The last
two mechanisms generate  photons  even in case of a vanishing
quark number, which is probably a good approximation for the early stage 
when the magnetic field is significant. 
The estimates given in Ref.~\cite{Basar:2012bp} showed that 
photon production from the conformal anomaly makes a considerable
contribution to $v_{2}^{\gamma }$ and  potentially can describe the measured data.  

\begin{figure}[t]
\includegraphics[width=8.0cm]{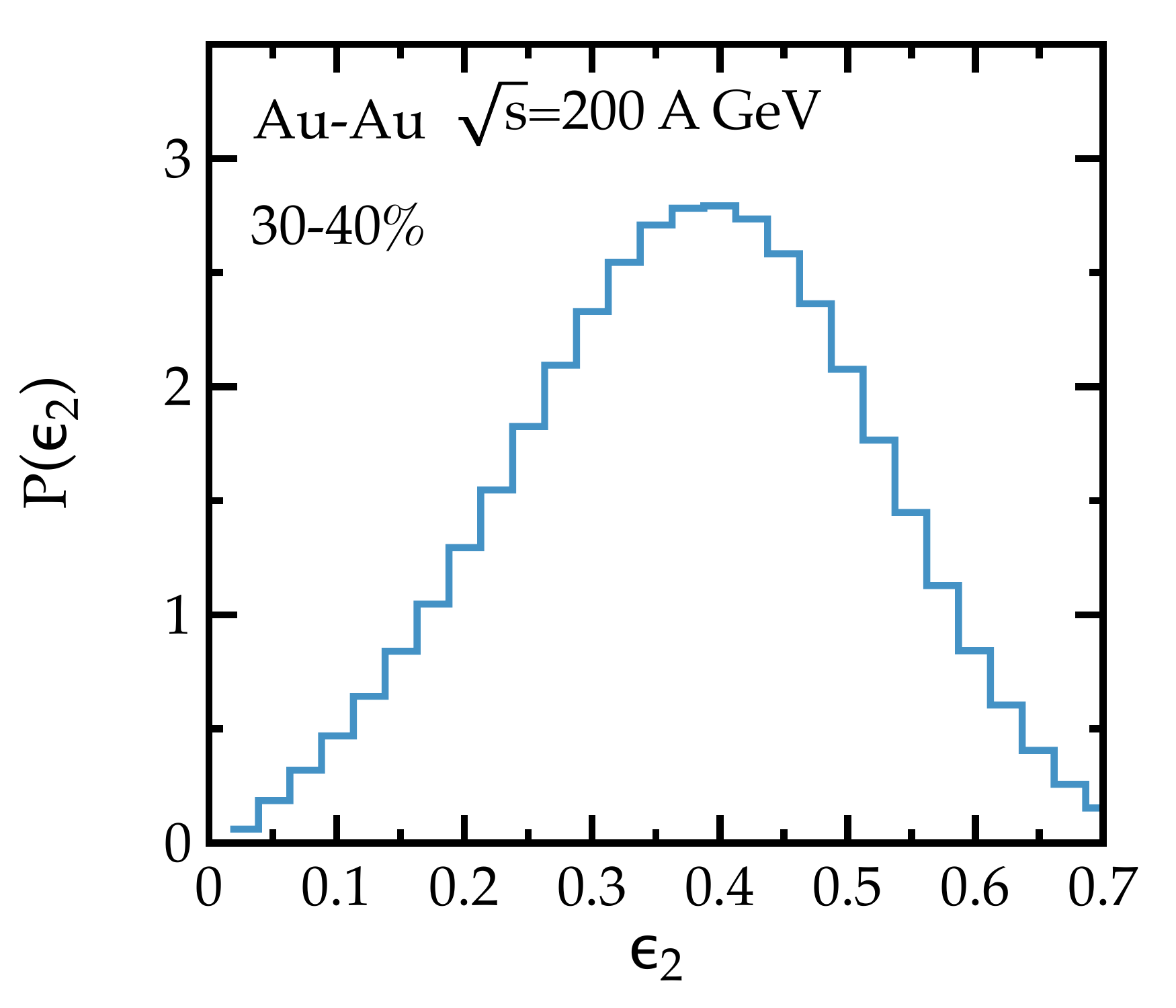}
\caption{ Probability distribution of the participant eccentricity $\epsilon_2$ for the centrality 
class 30-40\% in Au-Au collisions at $\sqrt{s}$=200 GeV. } \label{fig2}
\end{figure}

The question arises, as to how this could be experimentally establish if the
photons responsible for azimuthal asymmetry are produced from hadronic sources 
(or more generally from a non-zero eccentricity of the fireball), or from the quark-gluon 
plasma in the presence of a high magnetic field at early times. In this letter, we propose
an observable that can single out the mechanism responsible for
producing the azimuthal anisotropy of photons in heavy-ion collisions. We also
discuss other experimental signatures that may help to clarify the situation.

To proceed further, we first recall some properties of the
magnetic field in heavy-ion collisions. Firstly, the major contribution to
the magnetic field originates from the spectator protons of the colliding
nuclei. The charged particles produced may influence the magnetic field;
however, this effect is not expected to suffice   to overcome the field
of spectators at early times (see, e.g., transport model calculations in
Ref.~\cite{Voronyuk:2011jd}). Secondly, at a given centrality class the average magnetic field  almost
is independent of the fluctuating shape of the interaction region and the eccentricity $\epsilon_2$ of
the initial state; this problem was studied in more detail in Ref.~\cite{Bzdak:2011my}.

This feature will allow us to distinguish between different mechanisms of  $v_{2}^{\gamma }$ production. 
It is noteworthy that, in the first approximation, the eccentricity and the magnetic field 
are  linear growing functions of the impact parameter.   
Thus both mechanisms responsible for $v^{\gamma}_2$, hadronic flow and magnetic field, 
generate  approximately the same qualitative dependence of $v^\gamma_2$ on centrality. 
Consequently, this dependence cannot be used as a discriminative test of photon production. 

\begin{figure}[t]
\includegraphics[width=8.0cm]{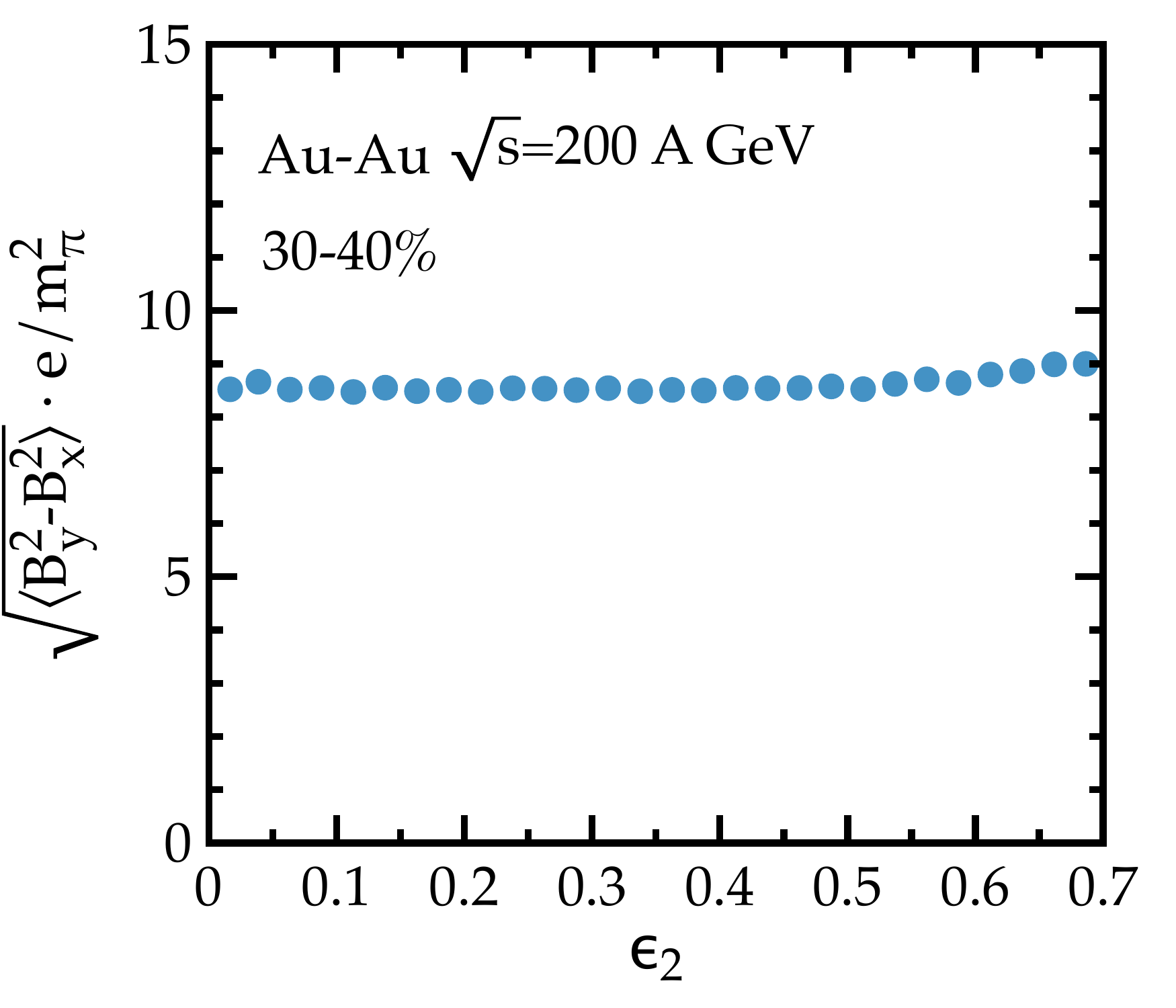}
\caption{ Dependence of the magnetic field asymmetry as a function of 
the initial eccentricity. $B_{y}$ and $B_{x}$ denote the out-of-plane and in-plane
components of the magnetic field. } \label{fig3}
\end{figure}

It is commonly accepted that hadronic flow in nucleus-nucleus collisions is defined by the
initial state's eccentricity, $\epsilon _{2}$. The
initial state's eccentricity is not only defined by the geometry of the
collision, but also by Glauber and ``intrinsic'' fluctuations (see e.g.~Refs.~\cite{Dumitru:2012yr,Schenke:2012hg}).
The first are related to the fluctuations of the nucleon positions in the
colliding nuclei, while the second represents  the fluctuations of the energy
deposition from interacting nucleons and their constituents. Consequently,
even in peripheral collisions the initial eccentricity $\epsilon _{2}$
strongly fluctuates leading to a broad range of 
values, as detailed in Ref.~\cite{Bzdak:2011my} where this problem was studied 
in the context of the chiral magnetic effect \cite{Kharzeev:2007jp}. 
In Fig.~\ref{fig2} we show the probability distribution of events as a function of the eccentricity $\epsilon_2$ 
for the centrality class $30-40\%$ in Au-Au collisions at $\sqrt{s}=200$ GeV. This result is obtained by 
using the standard Glauber 
model for  the initial state~\cite{Alver:2008aq}. As seen in Fig. \ref{fig2} the eccentricity distribution 
is quite wide, so we expect the same behavior for measured hadronic elliptic flow, denoted in this 
letter by $v^\pi_2$. 
In Fig.~\ref{fig3} the magnetic field asymmetry $\langle B_y^2-B_x^2\rangle ^\frac{1}{2}$, entering 
to the photon production rate of Ref. \cite{Basar:2012bp}, is shown. The calculations are performed by 
taking into account fluctuating proton positions in colliding nuclei. 
As seen in Fig.~\ref{fig3} the magnetic field is almost independent of the initial eccentricity 
for a given centrality class. We also checked that with a good precision, 
the average magnetic field $\langle B_y  \rangle$ coincides with $\langle B_y^2-B_x^2\rangle ^\frac{1}{2}$. 

Essentially, there are two competing mechanisms that can
contribute greatly to $v_{2}^{\gamma }$. The first ``hadronic''
mechanism~\cite{vanHees:2011vb} is related to the initial anisotropy of the
fireball~$\epsilon _{2}$: 
\begin{equation}
v_{2}^{\gamma }\propto \epsilon _{2}.
\label{hadronic}
\end{equation}%
The second mechanism reflects the presence of a strong magnetic field in the initial stage of a
collision, where %
\begin{equation}
v_{2}^{\gamma }\propto \left\langle B_{y}^{2}-B_{x}^{2}\right\rangle ,
\label{scale}
\end{equation}%
in the case of the scale anomaly of QCD$\times $QED~\cite{Basar:2012bp} and
the axial anomaly~\cite{Fukushima:2012fg}, and 
\begin{equation}
v_{2}^{\gamma }\propto \left\langle B_{y}\right\rangle ,
\label{synch}
\end{equation}%
for synchrotron radiation, as  discussed in Ref.~\cite{Tuchin:2012mf}. 
Here, 
$B_{y}$ and $B_{x}$ respectively denote the out-of-plane and in-plane
components of the magnetic field. 

\begin{figure}[b]
\includegraphics[width=8.0cm]{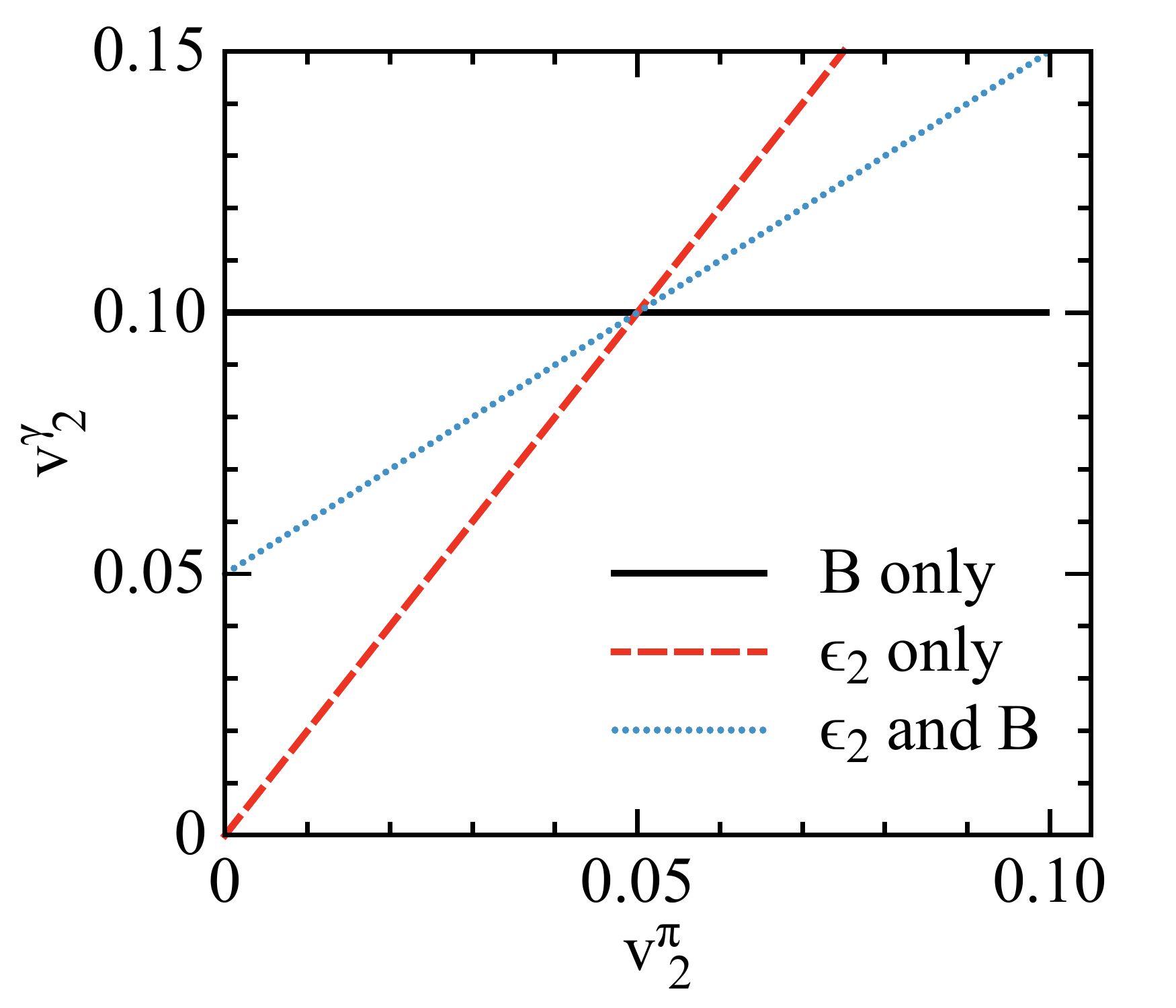}
\caption{Elliptic flow for photons $v_{2}^{\protect\gamma }(p_{t})$ at a
given $p_{t}$ as a function of the integrated elliptic flow for pions $%
v_{2}^{\protect\pi }$ for three possible scenarios. (i) Dashed (red) line: $%
v_{2}^{\protect\gamma }(p_{t})$ is dominated by the initial eccentricity $%
\protect\epsilon _{2}$. Here $v_{2}^{\protect\gamma }(p_{t})$ should be
proportional to $v_{2}^{\protect\pi }$ since $v_{2}^{\protect\pi }\propto 
\protect\epsilon _{2}$. (ii) Solid (black) line: $v_{2}^{\protect\gamma %
}(p_{t})$ is generated solely by the strong magnetic field. In this case $%
v_{2}^{\protect\gamma }(p_{t})$ should be approximately independent of $%
v_{2}^{\protect\pi }$ since the magnetic field at a given narrow centrality class
weakly depends on the fluctuating value of eccentricity $\protect\epsilon %
_{2}$. (iii) Dotted (blue) line: The case with equal contribution of $%
\protect\epsilon _{2}$- and $\vec{B}$-generated $v_{2}^{\protect\gamma %
}(p_{t})$. In this schematic plot we assume that, averaged over centrality
class \textrm{C}, $v_{2}^{\protect\pi }|_{\mathrm{C}}=0.05$, and $v_{2}^{%
\protect\gamma }(p_{t})|_{\mathrm{C}}=0.1$.}%
\label{fig1}
\end{figure}

Owing to eccentricity fluctuations at a given centrality class 
(or even at a given impact parameter), see Fig.~\ref{fig2},
we can select events with different values of the hadronic elliptic flow $v^\pi_2$, while centrality fixes 
the magnetic field, as we demonstrated in Fig.~\ref{fig3}. In the limiting case, viz., in non-central 
collisions of a certain centrality, 
we may select events with zero eccentricity and, consequently, with zero hadronic flow. 
This choice will eliminate contribution from Eq.~(\ref{hadronic}).  
If the magnetic field is responsible for the photon azimuthal anisotropy via Eq.~(\ref{scale}) or (\ref{synch}),
we still would  observe a non-zero $v_2^\gamma$ even for vanishing hadronic $v^\pi_2$.

Thus, the following three points summarize our idea to determine the mechanism of $v_{2}^{\gamma }$:

\begin{itemize}
\item First we choose a relatively narrow centrality class (e.g., $30-40\%$)
defined by the number of particles in the mid-rapidity region, or, if
possible, by measuring forward neutrons and protons.

\item Given the centrality class, in each event we measure the value of the
elliptic flow, $v_{2}^{\pi }$, for all pions in the hydrodynamic region (say $%
p_{t}<1$ GeV). It is commonly accepted that $v_{2}^{\pi }$ reflects the
initial eccentricity, $\epsilon _{2}$, of the fireball in the transverse
direction. Due to the fluctuations in positions of the participants, we obtain
a broad range of $\epsilon _{2}$ and consequently $v_{2}^{\pi }$.

\item Finally, we measure the elliptic flow for photons $v_{2}^{\gamma
}(p_{t})$ for different values of $v_{2}^{\pi }$. The magnetic field 
mainly is determined by the number of participants/spectators, and is rather
independent of the fluctuating values of $\epsilon _{2}$ in a given
narrow centrality class, see Fig.~\ref{fig3}. If $v_{2}^{\gamma }$ results solely
from the initial eccentricity then it should be proportional
to $v_{2}^{\pi }$. On the contrary, if the magnetic field dominates $v_{2}^{\gamma }$,
it should be independent of $v_{2}^{\pi }$.
\end{itemize}

Figure \ref{fig1} illustrates this idea, wherein we present three possible
situations: The photon anisotropy $v_{2}^{\gamma }$ is generated solely by the
initial anisotropy; $v_{2}^{\gamma }$ is generated by the magnetic field;
and  both mechanisms are present with
equal strengths.

Before concluding, several comments are warranted. The measurement discussed in 
this letter is best suited for mid-central and peripheral collisions, where both 
the elliptic flow and fluctuations of the initial eccentricity are expected to be
maximal. Also, the measurement should be performed for various values of photon transverse momenta. 
Possibly, different mechanisms of generating $v_{2}^{\gamma }$ may be applicable  
in different $p_{t}$ regions. Finally, the analysis should be performed in 
a narrow centrality class, e.g. $30-40\%$, so allowing  to neglect the correlation between $v_{2}^{\pi }$
and the impact parameter (and consequently, the value of the magnetic field) 
\footnote{In addition, the measurement of $v_{2}^{\pi}$ should be performed 
for a sufficiently broad rapidity window to ensure that the final  multiplicity
of hadrons and their azimuthal anisotropy adequately reflects the initial eccentricity, i.e.,
$\frac{1}{N_{\pi}} < v_{2}^{\pi}$, where $N_{\pi}$ is the number of hadrons 
used to determine $v_{2}^{\pi}$ }.

As was  suggested in Ref.~\cite{Basar:2012bp}, 
another probe for illuminating the mechanism of photon production 
lies in  the study of U-U collisions. The 
deformed shape of the U nucleus may allow to separate   
the eccentricity of the initial condition from the
magnitude of the magnetic field~\cite{Voloshin:2010ut}. 

Other crucial tests that can be performed to test the role of the magnetic field in heavy-ion
collisions include: 
\begin{itemize}
\item The violation of scaling $v_4/v_2^2 \sim 1$, which was observed for charged 
hadrons at PHENIX~\cite{Adare:2010ux}.
According to Ref.~\cite{Basar:2012bp} the anisotropic contribution to the 
photon production rate in a magnetic field 
is proportional to $k_x^2$ which only gives the second harmonics for azimuthal angle distribution.
Thus we expect $v^\gamma_4/(v^\gamma_2)^2 \ll 1$ for photons produced in a magnetic field.
\item While measurements of $v^\gamma_4$ require high statistics, it is sufficient to 
measure  $v^\gamma_3$, which in the first approximation is zero for photons produced in a magnetic field. 
\end{itemize}

In conclusion, 
we reiterate  that the recent PHENIX data on the photons' azimuthal anisotropy
raises new challenging  problems for the theoretical description of heavy-ion collisions.
This data either questions the conventional picture of early thermalization and subsequent
hydrodynamics of heavy ion collisions, or 
infers the existence of a new mechanism of photon production, i.e., contingent upon  
the magnetic field that can create
substantial photon azimuthal anisotropy. In this letter, we proposed 
a measurement of elliptic flow of photons $%
v_{2}^{\gamma }(p_{t})$ as a function of integrated elliptic flow of pions $%
v_{2}^{\pi }$, in a given narrow centrality class, e.g., $30-40\%$. 
This measurement will 
test both mechanisms of photon production due either  to the initial eccentricity
or the strong
magnetic field. We hope
that the proposed measurement would be useful both at RHIC and the LHC and, eventually, would allow 
to solve the puzzle of a large photon azimuthal anisotropy in heavy-ion collisions. 
We again stress that this may radically change our understanding of the 
non-equilibrium stage and thermalization in heavy-ion collisions.

\bigskip

\section{Acknowledgments}

We thank G. Basar, D. Kharzeev and L. McLerran  for discussions. 
We are grateful to A.~Woodhead for the careful reading of  the manuscript. 
The authors were supported by Contract No. DE-AC02-98CH10886 with the U.S. Department of Energy.
A.~B. also acknowledges the grant No. N202 125437 of the Polish Ministry of Science and
Higher Education (2009-2012).


\end{document}